\documentclass[universe,article,accept,pdftex,moreauthors]{Definitions/mdpi} 

\firstpage{1} 
\pubvolume{1}
\issuenum{1}
\articlenumber{0}
\pubyear{2025}
\copyrightyear{2025}
\datereceived{19 December 2024} 
\daterevised{21 January 2025} 
\dateaccepted{27 January 2025} 
\datepublished{ } 
\hreflink{https://doi.org/} 

\usepackage{soul}

\Title{Issues in the Investigations of the Dark Matter Phenomenon \mbox{in Galaxies}: {\it Parcere Personis, Dicere de Vitiis.}  }

\TitleCitation{Issues in the Investigations of the Dark Matter Phenomenon \mbox{in Galaxies}}

\Author{Paolo  Salucci} 

\AuthorNames{Paolo Salucci}

\AuthorCitation{Salucci, P.}

\address[1]{SISSA, Via Bonomea 265, 34136 Trieste, Italy. 
 salucci@sissa.it}
 
\abstract{It is always more evident that the kinematics of galaxies provide us with unique information on the  Nature of the dark particles and on the properties of the galaxy Dark Matter (DM) halos. However, in investigating this topic, we have to be very careful about certain issues related to the assumptions that we take or to the practices that we follow. Here, we critically discuss such issues, that, today, result of fundamental importance, in that we have realized that the  Nature of the DM will be not provided by ``The Theory'' but, has to be inferred by reverse engineering the observational scenario.}


\begin{document}


\section{Introduction}

 In the late 70', the kinematics of galaxies started to make very evident the presence in them of a non-luminous mass component (\cite{Rubin_1980}). In the following decades, these observations have increased in number, quality and importance and have set the scene for the Phenomenon of Dark Matter (DMP) at galactic scales ( \cite{Persic_1996}). However, it is well known that, today, observations do not comply (a 
 review in: \cite{Salucci_2019}) with the properties of the DM halo density profiles predicted by preferred theoretical cosmological scenarios, headed by the WIMPs, but including also Ultra Light Axions, Sterile Neutrinos, Self Interacting DM \mbox{(e.g., \cite{Hu_2000,Vogelsberger_2014}).} In this fundamental cosmological field we are now witnessing a change of paradigm: the strategy of solving the DMP mystery does not imply anymore to adopt a theoretical scenario for the dark particle and then to verify its predictions vs. observations score \cite{Salturini,Nesti_2023} but, instead, to exploit the observed properties of the DM in galaxies as a {\it portal} to the underlying theoretical scenario. Therefore, the galaxy kinematics takes a role much more relevant than that of a simple validation of some assumed physics, an then we must make sure that the process of study, analysis and interpretation of relevant data be extremely accurate and scenario-independent.
 
 Contrary to these requirements, several beliefs, assumptions and practices, well present in literature over the past 30 years, result incorrect/unjustified and jeopardize the investigation of the DMP. While the properties of the dark and luminous matter components in Galaxies have been determined in a number of works (see: \cite{Salucci_2019}), in the present one we aim to critically discuss such beliefs, assumptions and practices with the goal of clarifying the situation for the benefit of the scientific community. These issues are not ``niche topics'', but they concern fundamental aspects of the investigation of the DMP in galaxies. Moreover, we expect that this work will be of widespread interest in that a very large number of published works are involved \endnote{Number of involved works : > 55 k (ADS)}, at different levels, with the various issues that we will discuss in next sections. However, none of them will be cited in this work \endnote{\textit{Parcere personis, dicere de vitiis} (Valerio Marziale, {\bf 10}, 33, 10) } , in that our aim is just to point out, in the issues  
{\bf 1--6}, certain critical {\it situations  
 } in which the taken approach prevents the {\it body} of evidence obtained about the galaxy Dark Matter from {\it speaking} out about its  Nature.
 
 Let us also stress that this work concerns few specific aspects of the mass modeling of galaxies. A complete discussion of the pipeline required for the full investigation of the DMP in galaxies is beyond the scope of this work.

Let us summarize that the Phenomenon of Dark Matter at galactic scales includes the detection, in all objects, of large anomalous motions (see, e.g., \cite{Salucci_2019}): stars and gas inside them do not move as subjected to their own gravity, but as they were attracted also by another invisible component. Here, we will consider mainly disk systems i.e.,: spirals, dwarf irregulars and Low Surface Brightness, galaxies in which the RC is an accurate tracer of the total gravitational field and the surface photometry yields important information on the luminous disk component. For the disk systems, the equilibrium between the gravity force and the motions that oppose to it has a simple realization: the stars (and the HI gaseous disk) rotate around the galaxy center, so that: 
\begin{equation}
R \frac{d \Phi(R)}{dR} = V^2(R)
\end{equation}
where the (measured) circular velocity and the galaxy total gravitational potential: $V(R)$ and $\Phi(R)$ are related by the Poisson Equation. A disk of stars is their main luminous component, whose surface mass density $\Sigma_D(R)$, proportional to the surface luminosity measured by the photometry, takes the form of (\cite{Freeman}): 
\begin{equation}
\Sigma_\star(R)=\frac{M_{D}}{2 \pi R_D^{2}}\ e^{-R/R_{D}}
\end{equation}
with $M_D$ the mass of the stellar disk and $R_D$ its scale length measured from the photometry. With this distribution, one can take $R_{D}$ as the characteristic length-scale of the luminous component of these systems and the quantity $y \equiv R/R_D$ as the meaningful ``normalized'' radius that marks a specific position in the luminous galaxy, independently of the galaxy luminosity and surface brightness. Moreover, it is useful to define $R_{opt}\equiv 3.2 R_D$ the radius enclosing $83\%$ of the stellar disk mass, as the size of the stellar disk. \endnote{In this work we freely interchange the quantities $R_D$ and $R_{opt}\equiv 3.2 R_D$ as the ``size'' of the stellar disk. Furthermore, $V_{opt}\equiv V(R_{opt})$}

Being $V_{\star}(y)$ the luminous matter contribution to the circular velocity and by defining $v^2_\star(y)\equiv \frac{G^{-1}V_\star^2(y) R_D}{M_D}$ we arrive to the universal function ($I,K$ are the Bessel functions):
 \begin{equation}
 v_\star(y)^2=\frac{1}{2} \ y^2 (I_0\ K_0 -I_1\ K_1)_{y/2} 
 \end{equation}
 that specifies the contribution of the luminous matter to the circular velocity $V(R)$.

 In this work, we will deal in depth with a number of specific critical aspects of the investigation of the DMP.  Remarkably, the  outcome emerges to be of great importance   also for scenarios different from those defined by a  dark particle  and also able to better frame the observations at high redshift that are in tension with  the latter (e.g., \cite{Dia_2019,Jovanovic_2016, Genzel_2017,Jovanovic_2023,Mistele_2024}).

 Section \ref{s2} is dedicated to the implications for the DM component that arise from available kinematics while Section \ref{s3}, from those originated from how we model the galaxy DM halo
In Section \ref{s4} we discuss the results obtained.



\section{Issues with the Kinematics of Disk System}\label{s2}

An important and unexpected aspect of the rotation curve/circular velocity\endnote{In this work, we will freely exchange these two quantities} of disk systems is that their ``effective'' radial coordinate is not the physical radius $R$, but the normalized radius: $y \equiv R/R_D$ (e.g., see Figure 4 in \cite{Rubin_1985} and Figure 1 in \cite{Persic_1996}). This fact would be understandable if it was referred to the luminous component of the circular velocity of a galaxy, but, since it occurs also for the {\it sum in quadrature} of the dark and luminous components, becomes an open question of the galaxy mass distribution. One finds that an Universal relationship $ V_{URC}(y, V_{opt}$) well represents the great majority of the RCs: the two normalized circular velocities $V(R/R_{opt})/V_{opt}$ are well reproduced by a Universal Relation which extends from the galaxy center to the halo virial radius $R_{vir}$ (defined in Equation (9)), i.e., from $R=0$ to $R=R_{vir}\sim 50 \ R_D$ (see Fig (1)). In this work, we use the URC to represent the phenomenology of the kinematics of disk systems. No result of this work changes if, instead, we use an adequate ensemble of {\it individual} RCs. The building of such Universal Rotation Curve, its importance and extension to galaxies of different surface brightness and Hubble types are presented and discussed in: 
 \cite{Rubin_1985, Persic_1991,Persic_1996, Yegorova_2007, Salucci_2007,Catinella_2006,Karukes_2017,Lopez_2018,DiPaolo_2019}. 

I stress, then, that the existence of the URC implies that, in disk systems, the radial behavior of the structural properties of the dark and luminous matter components is framed by the quantity $R/R_D$ , rather than the galactocentric radius $R$. The rotation curve $V(y)$ of a disk system, with its dark and luminous components is given by: 
\begin{equation}
 V^2(y)=V^2_{DM}(y) +V^2_\star(y)
\end{equation}
 I also introduce two important kinematical quantities: $\nabla\equiv dlog\, V(y) /dlog \, y$, derived from $V(y)$ and $\nabla_\star \equiv dlog\, V_\star/dlog \, y$, known for the Freeman stellar disk with constant stellar mass-to-light ratio.\endnote{$\nabla_\star(y)\simeq 0.87 -0.5\ y +0.043\ y^2 $}

\begin{figure}[H]
\center
\includegraphics[width=0.77\columnwidth]{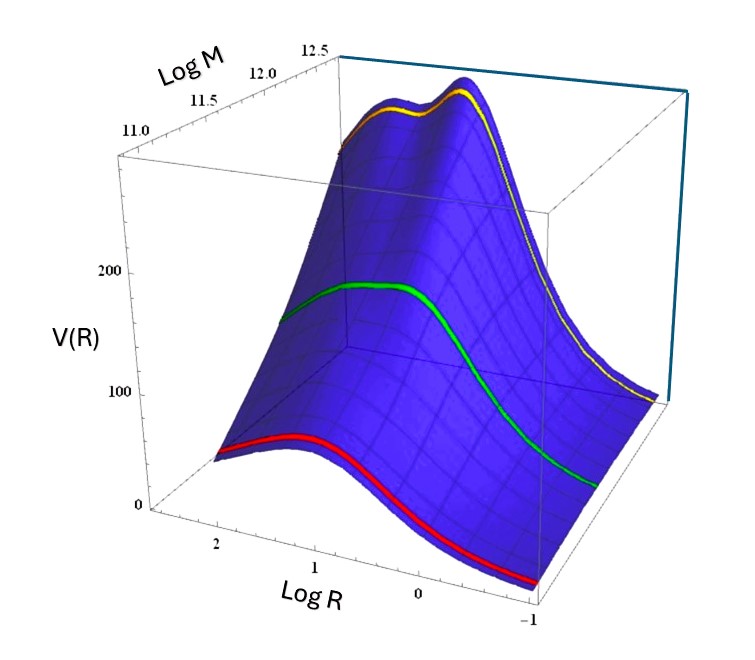}
\caption{The 
 Universal rotation curve of Spirals from $R= 0.1 \ kpc$ to $R_{vir}\sim  (100-300)\ kpc$  (see, e.g. \cite{Salucci_2007}).    $R$ is in kpc, $V$ in km/s, $M$ in solar masses . The ribbons indicate {\it (bottom-up) } the RCs corresponding to $M_{halo}\equiv M_{vir}=10^{11} M_\odot,10^{11.5} M_\odot, 10^{12} M_\odot$.  A simple visual impression rules out the claim of ``Flat Rotation Curves.''  \label{f1}}
\end{figure}

{\bf\large [1] } {\bf Flat Rotation Curves?} The RCs of disk systems cannot be considered as flat \endnote{Contrary to what it has been assumed in $> 8$\ k works (ADS)}, this condition, i.e.,: $\nabla(y)\simeq 0$, occurs rarely (see Figure \ref{f1}) and when it does, this has {\it very little} to do with the DM \endnote{Notice that in this issue we consider the RCs $V(y)$ in the region $0<y \leq 10-15 $ as done in the works that we critisize}. The RCs are generally described by locally rising or declining functions of radius (see Figure 1, \cite{Rubin_1985, Persic_1991,Persic_1996, Yegorova_2007, Salucci_2007,Catinella_2006,Karukes_2017,Lopez_2018,DiPaolo_2019}). In detail: in disk systems, from the center of the galaxy out to their virial radius, the RC slopes $\nabla(y)$ take all the possible values: from $1$, that of a rotating solid body, to $\sim -1/2$ that of a massless tracer in the Keplerian motion (e.g., see Figure 2 in \cite{Persic_1996}).

Furthermore, in disk systems, the condition $\nabla(y)\simeq 0$ has no special { \it physical meaning} in relation with the DM component: in fact, the RCs are flattish only in the following (little extended) regions: (a) in objects with: 50 km/s $\leq V_{opt} \leq $
 180 km/s, in the region: $4\,  R_D-6\,R_D$, (b) in objects with: 180 km/s $\leq V_{opt} \leq $
 300 km/s, in the region: $2 \, R_D- \, 4 R_D$ and (c) in objects with: 300 km/s $\leq V_{opt} $, in the region: $6 \, R_D- 8\, R_D$. (see Figure \ref{f1} and\mbox{ e.g., \cite{Salucci_2007}).} However, remarkably, in all these cases, the flattening of the RC is not caused by a {\it dominating} $R^{-2}$ {\it DM halo} as often believed, but, by a tuning between a {\it dominating stellar disk} contribution to $V(y)$ that decreases with $y$ and a {\it modest} halo contribution to $V(y)$ that increases with $y$. Furthermore, also when a RC is constant over a region, this is always followed by an external one in which the RC clearly increases or decreases, (e.g., \cite{Salucci_2007}). More specifically, the kinematical evidence for the presence of dark matter, is not provided by a {\it gentle flattening} of a moderately rising RC as often claimed, but by a {\it strongly rising} RC or by an {\it abrupt flattening} of a declining RC.

To assume that the Rotation Curves are flat in a large part of the region:  $0 \leq y\leq 15$ is, therefore, not supported by observations and it affects the investigation of the DMP not only by providing the centrifugal equilibrium condition of Equation (6) with a wrong quantity but, above all, because such assumption (wrongly) implies that: (a) the DM distribution is exactly the same in all objects and there is no structure parameter varying among them and (b) the DM density has an isothermal sphere profile whose enclosed mass is larger, at any radii, than that of the luminous component.  

 Then, in the region where we have kinematical data ($0 \lesssim  y \lesssim 15 $),  the idea of an asymptotically flat RC and that of the corresponding radius $R_{flat}$ from which such condition would start, are unsuitable: $R_{flat}$ does not exist in disk systems and, consequently $V(R_{flat})$ cannot represent a reference velocity for disk galaxies. Moreover, from the complex profile of the URC in Figure \ref{f1} we easily realize that $V(R_{last})$, i.e., the value of the rotation curve at the last measured point and $V(R_{max})$, the value at the radius $R_{max}$ at which the circular velocity has its maximum value,  are {\it misguiding} quantities: $V(R_{last})$ depends on the instrument employed to obtain the RC, while $R_{max}\sim 6 R_D$ for objects at low luminosities, but $R_{max}\sim R_D$ for objects of high luminosity (\cite{Salucci_2007}, a review in \cite {Salucci_2019}.)

 {\bf\large [2]} {\bf Investigating 
 the Dark Matter Phenomenon: does every Rotation Curve count? }

For a spherical dark halo one has: 
\begin{equation}
 V^2_{DM}(y)= 4 \pi G \ R_D ^2 \frac {\int\rho_{DM}(y) y^2 dy} {y} 
\end{equation}
where $\rho_{DM}(y)$ is the DM halo density. After some manipulations and by assuming, for simplicity, that the disk mass is obtained independently from the kinematics, one arrives to: 
\begin{equation}
 \rho_{DM}(y)=\frac{G^{-1}}{4 \pi R^2_D \ y^2}\Big(V^2(y)(1+2 \, \nabla(y))- V^2 _\star(y)(1+2 \,\nabla_\star(y)) \Big) 
 \end{equation}
where both $V_\star(y) = (0.2-1)\, V(y)$ and $ \nabla_\star(y)$ are known for a good number of objects via available mass models of individual or coadded RCs (e.g., \cite{Salucci_2019}). Let us notice that in Equation (6) the value of the DM density in the LHS emerges as a {\it fine-tuned difference} between the first and the second term of the RHS, both up to 10 times bigger than their difference. The kinematical 
 determination of the DM density, therefore, strongly depends on $V(y)$, $\nabla(y)$ and on their systematical and random errors. One can estimate that, a RC, in order to provide us with the mass distribution in a galaxy, must have intrinsic errors <10\% in $V(y)$ and not more than $\pm 0.15$ in $\nabla(y)$. If these conditions are not satisfied, the RC data trigger a totally wrong $\rho_{DM}(y)$ and a failure in the derived mass model. Noticeably,``bad '' RCs not only do not provide useful information, but lead us to totally wrong DM density profiles. This requires a proper selection of the RCs to use, a strategy not always put into practice, in that, although none claims to use bad quality RCs, however, in many works the rate of rejection of the available RCs is uncomfortably lower than that required. 

\section{Adopting A DM Halo Profile}\label{s3}

The main strategy to determine the DM halo density profile in disk systems is to model their circular velocity $V(y)$ with two components. One, $V_{bar}(y)$ includes all the baryonic matter, i.e., the stellar disk, the central bulge and the HI disk this can be obtained from the galaxy surface brightness, the stellar disk mass-to-light ratio and the HI surface density. Giving the goal of this paper, we consider only the Freeman stellar disk of measured length-scale $R_D$ and mass $M_D$, a simplifying, but not limiting assumption.
For the dark component, the spherical density profiles are usually adopted as result of one of the following approaches: (a) to use ``empirical'' profiles known to well perform on data: i.e., the Pseudo Isothermal (PI) and the Burkert (B) profiles \cite {Carignan_1985, Salucci_2000} or (2) to use ``theoretical'' profiles, i.e., the outcome of analytical computations or cosmological simulations performed within a specific Dark Particle scenarios e.g.,: the $\Lambda$CDM, ULA, SIDM ones. \cite{DiPaolo_2018_WDM,Hu_2000,Vogelsberger_2014,Boyarsky_2019}

Furthermore, one could adopt the ``general'' gNFW and Zhao profiles, i.e., empirical profiles that are claimed to be able to represent contemporaneously both the cuspy and cored profiles, by varying the values of their free parameters.

{\bf\large [3]} {\bf The 
 number of the free parameters of the Dark Matter halo density.} An important aspect of a DM density profile is the number of its free parameters: each profile, in fact, has a specific functional dependence of radius and a number of parameters $p_i$, whose scope is to take into account the halo by halo individualities of the DM density law, so: $\rho_{DM}=\rho_{DM} (R;p_1,p_2..)$. An incorrect choice in the number or  the  nature of these parameters is very likely able to jeopardize the ability of determining the dark halo density profile from the galaxy kinematics. 
For the empirical and the theoretical profiles, the number and values of the parameters are obtained by fitting the RCs or by means of other measures of the total gravitational field and from the outcome of cosmological simulations (or of analytical investigations), respectively. 

Remarkably, the observational profiles have been found to possess only {\it two} free parameters moderately related each other (\cite{Salucci_2019}). Similarly, the same number of free parameters, very often emerges for the theoretical ones \cite{Salucci_2019}. In detail, these quantities are: a DM halo density and DM halo length scale. Then, in disk systems and likely, in all galaxies, there is no reason to adopt halo profiles with more than 2 free parameters, until that emerges in some (new) DM particle or empirical scenarios. 

Moreover, Figure \ref{f1} with Equation (6) shows that, when expressed in physical units, the halo density of large galaxies $\rho_{DM} (R, LARGE)$ is different from that of the  small galaxies, $\rho_{DM} (R, SMALL)$.  An unique parameter-free functional form $\rho_{DM}=\rho_{DM} (R)$, able to account for all the DM halos around galaxies, does not exist. The DM halo density must have at least one free parameter, running (a lot) among galaxies. 

This could be problematic for theories alternative to DM with zero free-parameters, such MOND or for Modified Gravity frameworks with very constrained ``free parameters''.

{\bf\large [4]} {\bf Both 
 cored, but different.} To represent the DM halo density, often, one of the two following cored profiles is adopted, the Burkert profile (\cite{Salucci_2000}:

\begin{equation}
\rho_B (R)={\rho_0\, r_0^3 \over (R+r_0)\,(R^2+r_0^2)}~,\label{BH}
\end{equation}

or the PI profile (\cite{Carignan_1985}): 
\begin{equation}
\rho_{PI} (R)={\bar\rho_0\, \bar r_0^2 \over (R^2+\bar r_0^2)}~, 
\end{equation}
$\rho_0$, $r_0$, $\bar\rho_0$, $ \bar r_0$ are, respectively, the DM central density and the core radius for the two profiles. It is important to notice that, if for $R\leq r_0$, the two profiles are similar and interchangeable, for $R>r_0$ they strongly diverge. The PI profile, outside the region with available data, converge to a $R^{-2}$ behavior and keeps it out to the virial radius $R_{vir}$, the ``size'' of the DM halo, defined from the DM halo mass profile $M_{DM}(R)$ as:.
\begin{equation}
M_{DM}(R_{ vir})\equiv M_{vir}= 100~ 4/3\,\pi \ \rho_c \ R_{vir}^3
\end{equation}

The B profile, instead, in the same region, converges to the very well known collisionless NFW dark halo profile \cite{Navarro_1997}

\begin{equation}
\rho_{NFW}(R) = \frac{\rho_s}{(R/r_s)\left(1+R/r_s\right)^2},
\label{eq:nfw}
\end{equation}
where $r_{s} \simeq 8.8 \left(\frac{M_{\rm vir}}{10^{11}{M}_{\odot}}
\right)^{0.46}\ {\rm kpc}$ \cite{Klypin_2011} is the characteristic radius and $\rho_s$ the charecteristic density. Remarkably, for $R_{vir} \geq R \geq 1/5 \ R_{vir}$:

 \begin{equation}
 \rho_{NFW,B}(R) \propto R^{-(2.4 \pm 0.02)}.
 \end{equation}
different from the $R^{-2}$ behavior of the PI profile.
Then, according if we adopt a PI or a B fitting profile, the RC data at $R>r_0$ will contribute in a different way to the best fit of  \mbox{mass profile.}

One may adopt, for a cored distribution, the B profile considering that, in a dark particles scenario, it is difficult to imagine how they could maintain an isothermal $R^{-2}$ regime in the most external halo regions, given their extremely low densities of $(100-500) \rho_c$.
In any case, the PI profile cannot be adopted as the ``unique champion'' of the cored density profiles, as, instead, has occurred in a number of works \endnote{> 600 }.
 
 It is important to notice that   the outermost regions of the galaxies $$r_0<<0.2 \ R_{vir} \leq R \leq R_{vir} $$
  both for the {\it theoretical} NFW  and the {\it extrapolated} URC  scenarios, 
  the halo velocity ($\simeq V(R)$) decreases as $ \simeq (R/R_{vir})^{-0.2}$; however,  in this region, we do not have the {\it super-efficient} tracers of the gravitational field provided by the RCs and we have to resort   to model- dependent and complicated weak lensing analysis (e.g., \cite{Luo_2024}) that cannot determine the ``halo velocity'' $ \propto (M_{DM}(R)/R)^{1/2}$ accurately enough  to rule out the PI profile

{\bf\large [5]} {\bf General 
 Profiles for the DM halo density?} The gNFW \endnote{Adopted in $> 500$ works (ADS)} and Zhao (Z)\endnote{Adopted in $> 400$ works (ADS) } profiles (\cite{Zhao_1996} given by: 

\begin{equation}
\rho_{gNFW}(R)=\frac{\rho_{0}}{(\frac{R}{R_{0}})^{g}(1+(\frac{R}{R_{o}}))^ {3-g}} \,,
\label{eq:Zhao}
\end{equation}

\begin{equation}
\rho_{Z}(R)=\frac{\rho_{0}}{(\frac{R}{R_{0}})^{\gamma}(1+(\frac{R}{R_{o}})^{\alpha})^{\frac{\beta+\gamma}{\alpha}}} \,,
\label{eq:Zhao}
\end{equation}
where $\rho_{0}$ is the central density, $R_{0}$ the ``core radius'' and $g$, $\alpha,\beta,\gamma$ the slope parameters, have been proposed as ``all purpose'' halo profiles. By assuming them, one intends to remain in the comfort zone of the cuspy NFW profile, but being still able, if necessary, to go with the observational data leading to a cored configuration. 

Let us stress that the gNFW and Z $V_h(R)$ model, in successfully reproducing the RCs generated by NFW or B halos, must deal with the following nuisances: (i) the available data extend only over a limited portion of the whole RC, (ii) the intrinsically different functional form of the  B/NFW profiles with respect to the  Z/gNFW profiles (iii) the large number of parameters of the latter. On the top of this, the observational errors strongly affect the fitting procedure. The results, shown in Appendix \ref{sa}, clearly indicate the above nuisances determine the game: the gNFW and Z models, despite having one (three) extra free parameter(s) to play with, are unable to fit NFW or B halo velocities with moderate random errors. Then, in dealing with real data, they are useless: (a) in telling whether a RC favors a Burkert or a NFW halo, (b) in assessing whether a sample of galaxies contains objects with a NFW halo and others with a B halo, (c) in providing evidence that B halos originate from the NFW ones via some physical process and (d) in providing us with a cored halo profile which is minimally deviant from the NFW one. 

These two models, however, can be used if they emerge from some theory or out of cosmological simulations, or, if one chooses to adopt them as a new (exotic) empirical framework. However, in any case, they must be considered as: (a) new profiles to be compared with respect to any other proposed so far, and, (b) above all, completely different from the B and the NFW ones. 

Finally, it is worth to point out that to fit the RCs with a dark halo + luminous component mass model has other problems (discussed in literature) but outside the scope of this work.

\begin{figure}[H]
\center
\includegraphics[width=10cm]{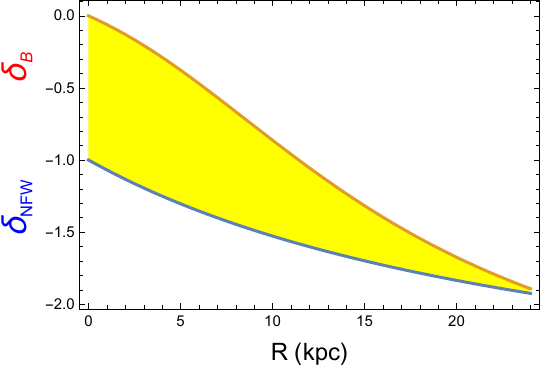}
\caption{The  
 slope of the DM density profile (Equation (14)) as function of radius for the NFW and B  halos of Eqs (A1), (A2) and Appendix \ref{sa}.  The yellow area indicates the difference between the  B ({\it orange}) and  the NFW ( {\it green}) slopes over the whole galaxy and it is inversely proportional to the cuspiness of the halo density. \label{g2}}
\end{figure}

 {\bf\large [6]} {\bf Cuspy 
 or cored DM halos?} In the past two decades the evidence that the DM halos have a cored density profile has been much debated in the literature (e.g., \cite{Bullock_2017,Del_Popolo_2009, Salucci_2001,Gentile_2004,Donato_2009}
 a recent review: \cite{Salucci_2019}). A decisive proof comes from the mass decomposition of the spiral's RCs in their dark and luminous components; however, only a small fraction of the available RCs are apt for this approach (e.g., see issue {\bf 2}). Then, a different method of analysis has been proposed that aims to exploit the different functional form of the two competing profiles. Let us focus on the quantity :
\begin{equation}
 \delta =\lim_{R\to 0} \frac{dlog\, \rho(R)}{dlog\,R}
\end{equation}
 it is then claimed that $\delta \simeq 0$ implies the presence of a cored DM distribution, while $\delta \simeq -1$, that of a NFW halo; values in between with their uncertainties, yield the probabilities of each of the two cases. Since, in order to estimate $\delta$, the requirements on the RCs and the assumptions in the distribution of the luminous matter are less severe than those needed for a full mass modeling, this approach has been appreciated and used several times. 
 However, there is one point to consider: from the RC mass modeling we realize that, while the cored region of the B profile extends out to $r_0\sim 3 R_D$, the DM density is not a {\bf constant 
} function of radius from $R=0$ to $R=r_0$, where, in fact, its value is only $1/4$ that of the central one (see Equation (7)). $\delta_B$ is then about zero only very near the center and from there it starts to rapidly decrease with radius (see Figure \ref{g2}). Also in the NFW framework, in which $r_s \sim r_0 $, we have a similar instance: the value $\delta_{NFW} = -1$ occurs only in the galaxy center and from there outwards, this quantity starts to rapidly decrease with radius (see Figure \ref{g2}). 
 
 It is important to stress that in the available RCs the slope $\delta$ cannot be measured at $R=0$, the innermost radius where this can be safely done is at: $R\simeq (1-1.5) R_D$. Thus, the expected score of the game: NFW {\it vs} B is not: $-$1 {\it vs} 0, as often assumed, but something like: $-$1.4 {\it vs} $-$0.7. Therefore, also values of $\delta$< $-$0.5 may indicate a cored profile and the expected value of $\delta_{NFW} $ is $< -1$.

\section{Conclusions }\label{s4}

This aim of this work was not to review the fundamental properties of the dark and the luminous matter in galaxies (as in, e.g., \cite{Salucci_2019}), but to help the researchers on this topic to avoid certain assumptions and practices that could jeopardize their investigation by preventing the available ``body of observations'' from {\it talking} and telling us the  Nature of the dark component. In other words, the kinematics of disk systems can provide us with an unique information about their dark and luminous matter components. However, if such information is not retrieved in the correct way, the outcome on the DM  Nature and its astrophysical role will be subject to errors and biases. That is, this work suggests that Cosmologists should consider the possibility ``to renew their toolboxes'' with which, by reverse engineering observational data, they address the Physics of Dark Matter \mbox{in Galaxies.}


\funding{This research received no external funding. 
}

\dataavailability{Data are available upon request  
}

\conflictsofinterest{The authors declare no conflicts of interest.}

.
\appendixstart
\appendix 
\section[\appendixname~\thesection]{}\label{sa}

Within the NFW and B scenarios we build two reference halo velocity models \cite{Salucci_2007}:
 
\begin{equation}
V_{\rm NFW}^2(R)=V_{\rm vir}^2 \frac{1}{X} \frac{{{\ln}} (1 + c R) - \frac{c R}{1 + c R}} {{{\ln}} (1 + c) -\frac{c}{1+c}}\,,
\end{equation}
\begin{equation}
 V^2_{B} (R) = \frac{G}{R} 2\pi\rho_0r_0^{3} \left[\ln(1+R/r_0) + \frac{1}{2} \ln (1+R^2/r_0^2)- \tan^{-1} (R/r_0) \,.\right]
\label{eq:Burk}
\end{equation}
in both cases, we take a reference mass of $M_{vir}=2 \times 10^{12} M_\odot$ with the related virial radius: $R_{vir}=260 \ kpc$. For the B halo we take: $\rho_0 = 4.7 \ 10^{-25}$ \ g/cm$^3$, so that: $r_0 = 18 \, kpc $, while for the NFW halo we take: $ c=R_{vir} / r_s = 9.0 $ with 
$M_{vir} = 200 \rho_c \ R_{vir}^3=G^{-1} V_{vir}^2 R_{vir}$. In order to mock actual RCs we set the extension of the above RCs from $2\ kpc$ to $50 \ kpc $. \endnote{The results in this section do not depend on these assumptions.}

\begin{figure}[H]

\begin{adjustwidth}{-\extralength}{0cm}
\centering 
\includegraphics[width=0.55\columnwidth]{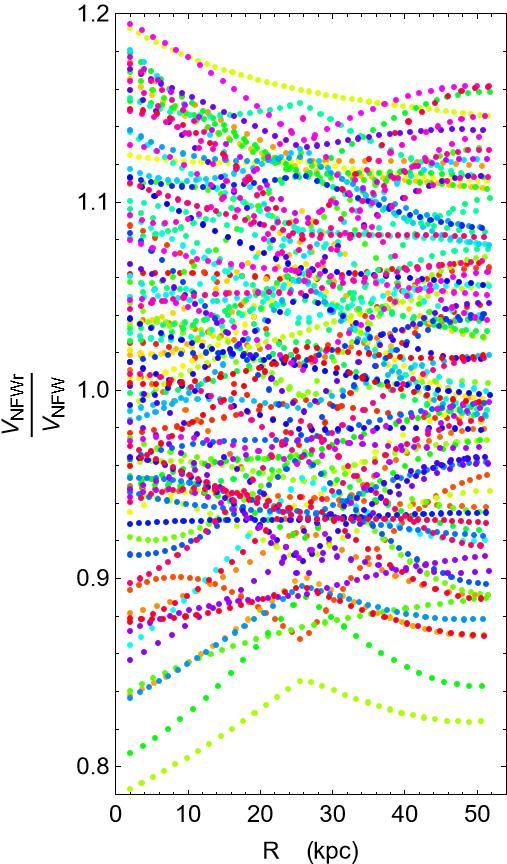} 
\includegraphics[width=0.55\columnwidth]{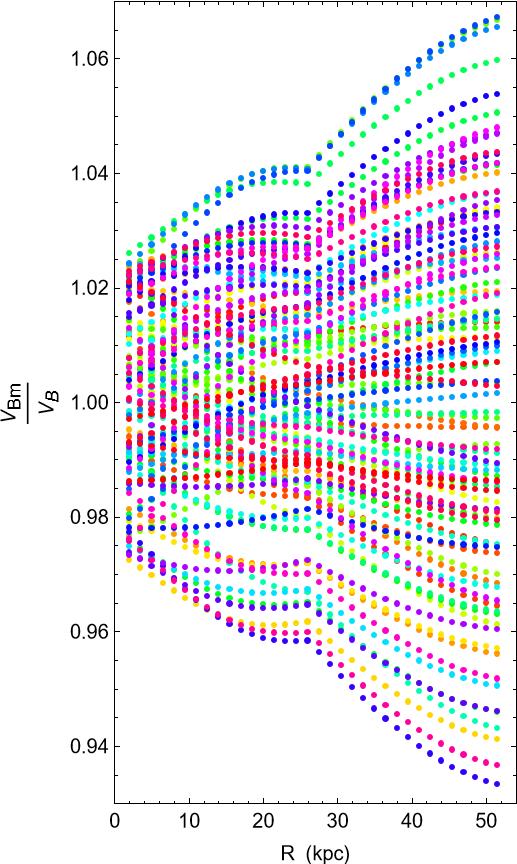}
\end{adjustwidth}

\caption{100 NFW (\textbf{left}) and 100 B (\textbf{right}) mock halo velocities as function of radius, generated from the reference velocities in Equations (A1) and (A2) as explained in Appendix \ref{sa}. The mock velocities are expressed in units of the corresponding NFW/B reference halo velocities. \label{A1}}
\end{figure}

Then, by inserting in Equations (A1) and (A2) the above values of the two pairs $(c,V_{vir}) $\ and $(\rho_0,r_0)$ we obtain the NFW/B reference halo velocity models. By convolving these two models with moderate random errors, we create 100 B and 100 NFW DM halo velocity components of the circular velocity hereafter, just halo velocities or $V_h$ (see Figure \ref{A1}). In detail, we substitute, in the two reference halo velocities, the values of $(c,V_{vir})$ and $(\rho_0,r_0)$ with two pairs of mock values: ($c_m, V_{virm})$, $(\rho_{0m},r_{0m})$ where:
\begin{equation}
 c_m = c\, (1 + RR[-0.15, 0.15]) 
\end{equation} 
$$ 
V_{virm}= V_{vir}\, (1 + RR[-0.12, 0.12])(1 + RR[-0.08,0.08]) \, | \, cos(5 \,\pi\, R/R_{vir}]|)$$
$$
\rho_{0m} = \rho_0 \, (1+(1+0.3 |\ cos(5\, \pi \, R/R_{vir})|) \,RR[-0.4,0.4])
$$
$$r_{0m} =r_0\, (1+ RR[-0.04,0.04])$$

$RR[x_1,x_2]$ is a random value between $x_1$ and $x_2$ that introduces a random fluctuation in the structural parameters of the halo velocities, while the cosine terms adds a random fluctuation in the halo velocities at their innermost and outermost radii $(2\, kpc, 52 \, kpc)$, taking into account that, in these regions, data are more uncertain. In short, we have moderately randomly perturbed the amplitude and the slopes NFW/B halo velocities. We investigate whether the 100 $V_{NFWm}(R)$ and 100 $ V_{Bm}(R)$ mock halo velocities, shown in Figure \ref{A1} normalized to the reference velocities, can be reproduced by the gNFW and Z models. For simplicity, we consider the disk component in the circular velocity as known, so that we can neglect them in the analysis.

In the gNFW case, the fitting halo velocity (with 3 free parameters: $g, c,M_{vir}$) reads as:
 \begin{equation}
V_{gNFW}(R;M_{vir},c,g) = 2081 R^{-1/2} A(R; M_{vir},c,g)/B(R;c,g)
 \end{equation}
$$
A(R;M_{vir},c,g)= C(R,M_{vir})
\Big(\int_{0}^{1}(y^2\, (y\, c\, R /R_{vir})^{-g} (1 + c \, y \,R/R_{vir})^{-3 +g} \, dy\Big)^{1/2}
$$
 $$
 B(R;c,g)= \Big(\int_0^1(z^2 (c\, z)^{-g} (1 + c\, z)^{-3 + g} \, dz\Big)^{1/2} \ \ \ \ C(R,M_{vir})=\Big(\frac {R}{R_{vir}}\Big)^{3/2} \Big(\frac {M_{vir}}{10^{12}M_\odot}\Big)^{1/2}
$$
where $R_{vir}, R $ are in kpc.

In the Z case, the fitting halo velocity, with 5 free parameters $ \rho_s,r_s,\alpha, \beta,\gamma$ (a reference density, a reference radius and three slopes of the density profile), reads as: 
\begin{equation} 
V_Z(r;\rho_s,r_s,\alpha, \beta,\gamma) = \Big(D(r;r_s,\gamma) \, _2F_1\Big[\frac{3 - \gamma}{\alpha}, \frac{\beta + \gamma}{\alpha}, 1 + \frac{3 - \gamma}{\alpha}, -(\frac{r}{r_s})^\alpha\Big]\Big)^{1/2}
 \end{equation}
 $$
D(r;r_s,\gamma) =28.2 ^2\, (3 - \gamma) ^{-1}\rho_s r^2 (r/r_s)^{-\gamma} \ (km/s)^2
$$
where $ _2F_1$ is the Hyper-geometric function, the radius is in unit of kpc, $\rho_s$ in unit of $10^{-22}$ g/cm$^3$.

The first step is to fit, separately, the two ensembles of mock velocities, each one with its generating NFW/B model with its two free parameters. Not unexpectedly, these velocity models fit well the generated mock velocities and recover, within reasonable uncertainties, the values of the parameters set for the reference models (see Figure \ref{A2}). Instead, $V_{gNFW} $, generally fails to reproduce the mock NFW and B halo velocities (see Figure \ref{A2}) and the resulting best-fit free parameters are in serious discrepancy with those of the reference halo velocities. $V_Z$ performs even worse (see Figure \ref{A2}) in fitting the NFW/B mock velocities. Furthermore, the best-fit values of the parameters of the Z model are found to vary among the sample of mock velocities by a factor 5--10, despite the fact that all of them originate from a same velocity profile plus a modest random perturbation.

\begin{figure}[H]
\includegraphics[width=0.6\textwidth]{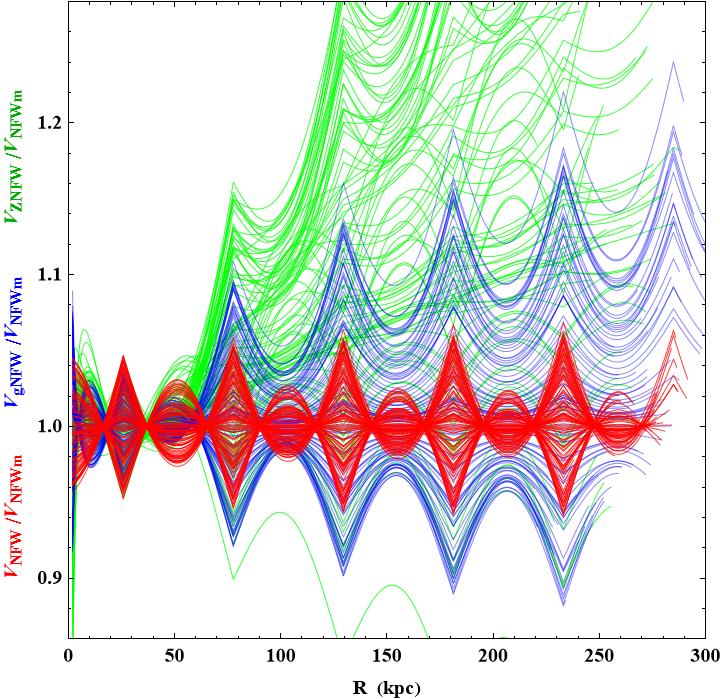}
\caption{\textit{Cont}. \label{A2}}
\end{figure}

\begin{figure}[H]\ContinuedFloat
\includegraphics[width=0.71\textwidth]{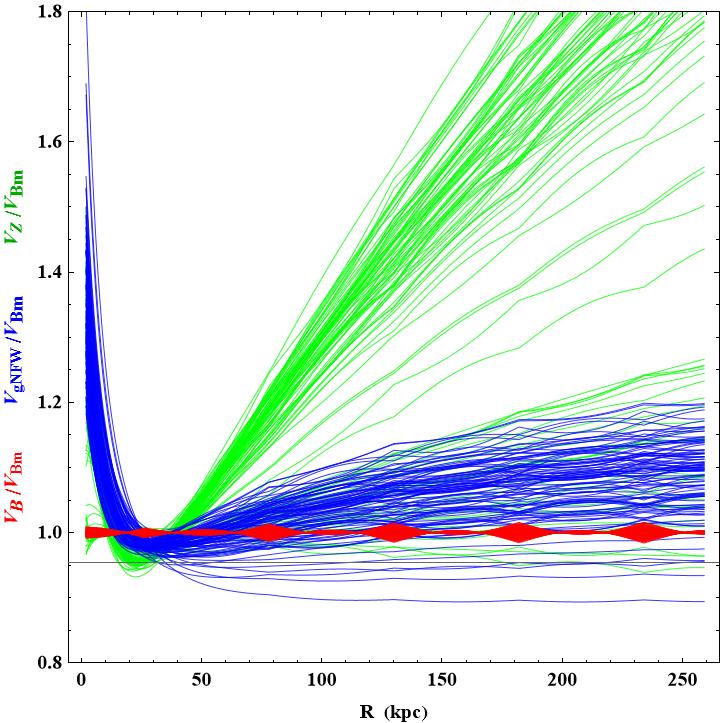}
\caption{100 $\times$ 3 best fit halo velocities as function of radius, expressed in units of their corresponding Burkert mock velocity. (Red, Blue,Green) \ color = (Burkert, gNFW, Z) \ profile. \label{A2}}
\end{figure}

.

\begin{adjustwidth}{-\extralength}{0cm}

\printendnotes 

\reftitle{References}

\PublishersNote{}
\end{adjustwidth}
\end{document}